\documentclass[12pt]{article}
\usepackage{authblk}
\usepackage{graphicx}
\usepackage{cite}
\usepackage{amssymb,amsmath}
\usepackage{multirow}
\usepackage{footnote}
\usepackage[para]{threeparttable}
\begin{document}
\title{Interaction of hydrogen peroxide molecules with non-specific DNA recognition sites}%
\author{D.V.~Piatnytskyi}
\author{O.O.~Zdorevskyi}
\author{S.N.~Volkov}

\affil{Bogolyubov Institute for Theoretical Physics of the National Academy of Sciences of Ukraine, 14b, Metrolohichna Str., Kyiv 03143, Ukraine, snvolkov@bitp.kiev.ua}

\setcounter{page}{1}%
\maketitle

\begin{abstract}
Ion beam therapy is one of the most progressive methods in cancer treatment. Studies of {the} water radiolysis process show that the most long-living species that occur in the medium of {a} biological cell under the action of ionizing irradiation are hydrogen peroxide (H$_2$O$_2$) molecules. But the role of {H$_2$O$_2$} molecules in the DNA deactivation of cancer cells in ion beam therapy has not been determined yet. In the present paper{, the} competitive interaction of hydrogen peroxide and water molecules with atomic groups of non-specific DNA recognition sites (phosphate groups PO$_4$) is investigated. {The} interaction energies and optimized spatial configurations of the considered molecular complexes are calculated with the help of {molecular mechanics} method and quantum chemistry approach. The results show that {the H$_2$O$_2$} molecule can form a complex with {the} PO$_4$ group (with and {without a} sodium counterion) that is {more} energetically stable than the same complex with {the} water molecule. Formation of such complexes can block genetic information transfer processes in cancer cells and can be an important factor during ion beam therapy treatment.
\end{abstract}

\section{Introduction}
\label{intro}

Ion beam therapy is one of the most {prospective} methods of {radiation therapy}. {To treat patients, it} uses heavy ion beams {(usually protons and carbon ions)} produced on special accelerators~\cite{Kraft2000,durante2019proton}. {Ion beam therapy is based on} the so-called Bragg's effect~\cite{bragg1904}, when {the beam linear energy density has the maximum at the end of the ions' track}. Due to this effect, a cancer tumour can be destroyed without {any} significant influence on healthy tissues.

{Molecular} radiobiology assumes~{\cite{Sarkaria2008}} that {in order} to destroy {a} cancer cell its DNA must be deactivated. The most powerful mechanisms of this deactivation are DNA double-strand breaks~\cite{khanna2001dna,schipler2013dna}. According to the multiscale approach of ion beam therapy~\cite{surdutovich2009ion,solovyov2010}, {the} double-strand breaks are caused by secondary electrons, free radicals, or by heating of {the} intracellular medi\-um. {Despite an important role that DNA strand breaks play in radiation therapy,} the {repair} mechanisms that {can deal with the strand} breaks~{\cite{NobelPrize2015,Stover2016,biau2019altering}} are present in {a} living cell. {Therefore, it is worthwhile to look for some other DNA deactivation mechanisms that can take place during ion beam therapy and can be less affected by the repair apparatus of the cell.}

Due to {a} large amount of energy transferred from the beam to {the} water medium {of a cell}, the water radiolysis process takes place. As a result, in the Bragg peak area, a wide variety of species occur, such as secondary electrons, free radicals, ions as well as molecular products ({hydrogen peroxide and hydrogen molecules}). {According to the  works~\cite{shockwave2010,solovyov2018}, chemically reactive species can spread over significant distances from the ion track due to the formation of a shock wave around the pathway of the beam.} Monte Carlo simulations~\cite{Durante2018,Kreipl2008,Uehara2006} revealed that on the time scales of the physiological processes ($\sim 1$ $\mu s$) the highest concentration have hydrogen peroxide (H$_2$O$_2$) molecules. This takes place because the free radicals recombine on much shorter time scales (chemical stage of water radiolysis)~\cite{leCaer2011}. {Moreover, H$_2$O$_2$ molecules can live in water medium for longer times (more than $\sim$ 1 $\mu s$) and diffuse for significant distances (more than $\sim$ 200 nm) from the track of an incident particle~\cite{Durante2018}}. Therefore, {H$_2$O$_2$} molecules have an {enhanced} probability to {\textquoteleft}find{\textquoteright} the DNA molecule in the intracellular water medium.

{Up to now, no enough essential attention has been paid} to the role of hydrogen peroxide in the context of ion beam therapy. But its participation in other methods of cancer treatment has been {discussed already} in the literature. In this way, the work~\cite{levine2013} emphasized the significant role of hydrogen peroxide in the cancer treatment by ascorbic acid. The experiment in~\cite{chen2005} showed that hydrogen peroxide causes more damage to cancer cells whereas healthy cells are less vulnerable to its action. {However, the certain molecular mechanism that lays behind this result has not been considered.}

In our work~\cite{pyatEPJ2015} the mechanism of the DNA deactivation by hydrogen peroxide was proposed. According to our hypothesis, H$_2$O$_2$ molecules (that occur in the medium of the cell due to water radiolysis) create stable complexes with the DNA atomic groups and in such way can block the processes of the DNA recognition by {an} enzyme. Consequently, the atomic groups become no more {recognizable} by the enzyme and further processes of cell division in cancer cells will be suspended.

To understand the complicated molecular mechanisms that take place in living cells, {it will be useful to analyze the interactions of H$_2$O$_2$ molecules with the important atomic groups of the double helix (recognition sites).} These sites {can be} divided {into} non-specific (the {phosphate PO$_4^-$ groups} of the DNA backbone) and specific (DNA nucleic bases)~\cite{Saenger}.  As hydrogen peroxide and water molecules have {a} similar structure, they can compete for binding with the DNA active sites. The competitive interaction of {H$_2$O$_2$ and H$_2$O} molecules with specific DNA recognition sites has been {studied already} in our previous works~\cite{ZdorevskyiDopovidi2019,ZdorevskyiUjp2019}. The result has shown that {an H$_2$O$_2$} molecule can bind to the atomic groups of the individual bases stronger than a water molecule. At the moment an accurate analysis of the interactions of H$_2$O$_2$ molecules with non-specific DNA recognition sites is required. 

The interactions of {a} hydrogen peroxide mole\-cule with {a} DNA phosphate group was studied in our work~\cite{pyatEPJ2015}. Using {molecular mechanics method}, the interaction energy in the considered complexes was calculated. It was shown that hydrogen peroxide {can} form a complex with PO$_4^-$ group which was no less stable than the same complex with {a} water molecule. Calculations performed in the mentioned work were made in {a} vacuum. {This} led to the anomalously large interaction energies in the framework of {the molecular mechanics method}~\cite{pyatEPJ2015}.

In the present work{,} the interaction of H$_2$O$_2$ and H$_2$O molecules with DNA PO$_4^-$ groups will be considered {by} taking {the implicit solvent into account}. {The} solvent {will simulate} the interaction of DNA atomic groups with the water medium in {a} living cell.

The goal of the present work is to determine the stable complexes which consist of hydrogen peroxide and DNA phosphate group in the presence of sodium counterion and to analyse the possibility of the blocking of non-specific DNA recognition sites by hydrogen peroxide. In Sec. \ref{methods} our calculation methods are described.  In Sec. \ref{results} different complexes consisting of {a} PO$_4^-$ group, sodium counterion, hydrogen peroxide and water molecules are considered. Using {molecular mechanics} and quantum chemistry approach, interaction energies of the considered complexes are calculated. {In Sec. \ref{blocking}}, the possibility of blocking DNA genetical activity by hydrogen peroxide molecules is discussed.

\section{Calculation methods}
\label{methods}
For the analysis of interaction energy and structure of the investigated molecular complexes two computational approaches are used - the {molecular mechanics (MM) method} and the method of quantum-chemical calculations on different levels of theory.

\subsection{{Molecular mechanics}}
\label{MM}
The method {of molecular mechanics} is now widely used in the modern molecular dynamics force fields~\cite{Charmm,Amber,LaveryMolDynRev} for studying the structure of molecular complexes. In the framework of this method, the energy of intermolecular interaction consists of hydrogen bonds ($E_{HB}$), van der Waals ($E_{vdW}$) and Coulomb ($E_{Coul}$) interactions:
\begin{equation}
E(r)=\sum_{i,j}(E_{HB}(r_{ij})+E_{vdW}(r_{ij})+ E_{Coul}(r_{ij})).
\end{equation}

Van der Waals's interaction between the {atoms $i$ and $j$ separated by the} distance of $r_{ij}$ is described by {the} Lennard-Jones {\textquoteleft}6-12{\textquoteright} potential:
\begin{equation}
E_{\mathit{vdW}}\left(r_{\mathit{ij}}\right)=-\frac{A_{\mathit{ij}}}{r_{\mathit{ij}}^6}+\frac{B_{\mathit{ij}}}{r_{\mathit{ij}}^{12}},
\label{for:vdW}
\end{equation}
where the parameters $A_{ij}$, $B_{ij}$ are taken from the work~\cite{Poltev1980} {(Tabl. S1, Supp. Mat.)}.

{As the hydrogen bond interactions are about an order of magnitude stronger than the van der Waals ones~\cite{Saenger}, the interaction energy between the atoms that form H-bond} is modelled by the modified {Lennard}-Jones potential {\textquoteleft}10-12{\textquoteright}:

\begin{equation}
E_{\mathit{HB}}\left(r_{\mathit{ij}}\right)=\left[-\frac{A_{\mathit{ij}}^{\left(10\right)}}{r_{\mathit{ij}}^{10}}+\frac{B_{\mathit{ij}}^{\left(10\right)}}{r_{\mathit{ij}}^{12}}\right]\cos \varphi_{ij} ,
\label{for:HB}
\end{equation}
{which is much sharper and deeper than the `6-12' potential (\ref{for:vdW}), see Fig. S1 (Supp. Mat.). Here $r_{ij}$ is the distance between the atoms i and j;} $A_{ij}^{(10)}$, $B_{ij}^{(10)}$ are the parameters taken from~\cite{PoltevShul1986} {(Tabl. S1, Supp. Mat.). This potential also takes into account the H-bond bending angle ($\varphi_{ij}$).} For example, when the hydrogen bond is $O-H ... N$, then  $\varphi $  is an angle between the lines of {the} covalent bond ($O-H$) and the hydrogen bond ($H ... N$). {Introduction of $cos \varphi_{ij}$ into {the} hydrogen bond potential (\ref{for:HB}) was made in~\cite{lavery1986flexibility}. {This} allows to account the weakening of H-bond interactions with the increase of a bending angle.}

Coulomb interaction is described by the electrostatic potential:
\begin{equation}
E_{\mathit{Coul}}\left(r_{\mathit{ij}}\right)=\frac 1{4\pi \varepsilon _0\varepsilon
    \left(r_{\mathit{ij}}\right)}\frac{q_iq_j}{r_{\mathit{ij}}},
\label{for:Coulomb}
\end{equation}
where $q_i$ and $q_j$ are the charges of the atoms $i$ and $j$ located at a distance $r_{ij}$, $\varepsilon_0$ is the vacuum permittivity, and $\varepsilon(r)$ is the dielectric permittivity of the medium.

Charges on the atoms of water and hydrogen peroxide molecules are calculated from the condition that the dipole moment of H$_2$O molecule should be equal to $d_{H2O}=1.86$ $D$~\cite{clough1973dipole}, and of H$_2$O$_2$ molecule $d_{H2O2} = 2.10$ $D$~\cite{massey1954microwave}. {The calculated charge values are listed in Tabl. S1 (Supp. Mat.). The obtained charges for {the} hydrogen peroxide molecule } are in good agreement with those obtained in the previous quantum-chemi\-cal calculations~\cite{Moin2012} {and are used in the recently developed force field for hydrogen peroxide~\cite{orabi2018}.}

The interaction of the charged oxygen atoms {of PO$_4^-$} with so\-dium counterion is {modelled} by the Born-Mayer potential~\cite{kittel1996} that takes into account the repulsion of atoms {at} short distances:
\begin{equation}
\label{for:bornMayer}
E_{BM}(r_{ij})=E_{Coul}(r_{ij})\lbrack 1-\frac{br_{ij}}{r_0^2} exp(-\frac{r_{ij}-r_0}{b})\rbrack,
\end{equation}
where $b=0.3$ {\AA} is the repulsion constant and $r_0=2.35$ {\AA} is the equilibrium length. These parameters are taken from works~\cite{perepelytsya2004UJP,perepelytsya2007counterion}.

{A} more effective accounting of Coulomb interactions can be achieved using the dependence of the dielectric permittivity upon distance ($\varepsilon $(r)). {It was derived} by Hingerty \textit{et al.}~\cite{hingerty} in the form:

\begin{equation}
\label{for:hing}
\epsilon \left(r\right)=78-77\left(r_p\right)^2\frac{e^{r_p}}{\left(e^{r_p}-1\right)^2},
\end{equation}
where  $r_p=r/2.5.$  Accordingly, the {molecular-mecha\-nics} method ({MM}) with the use of expression (\ref{for:hing}) will be marked as {MMh}. The corresponding calculations in {a} vacuum will be called {MMv}.

{In the framework of the present approach, we consider all the covalent bonds and angles as rigid. {This is done in order to} understand the role of the flexibility of a hydrogen peroxide molecule when comparing the results of the present method to the quantum-chemical approach, where all the geometries of individual molecules are flexible. Particularly, the effect of the flexibility of the dihedral angle of H$_2$O$_2$ on the interaction energy values is of main interest.} {The} geometries of the individual molecules that are considered for calculations in {the MM} method are presented in our work~\cite{pyatEPJ2015}.

\subsection{Quantum-chemical approach}

In the framework of quantum-chemical approach, the Hart\-ree-Fock method (HF/6-311+G(d,p)), density functional theory
(B3LYP/6-311+G(d,p)) and Moller-Ples\-set perturbation theory (MP2/6-311+G(d,p)) within the Gaussian 03~\cite{gaussian} program are used. {The} calculations are performed for complexes in {the} gas phase and water solution. To take {the} implicit solvent into account, the polarizable continuum model (PCM) of water solution is used. Geometries of H$_2$O$_2$ and H$_2$O molecules as well as of the phosphate group PO$_4^-$ are optimized within each of the methods. Interaction energies ($\Delta E$) in the considered molecular complexes are calculated using {the} supermolecular approach. Within this approach, $\Delta E$ is defined as the difference between the total energy ($E$) of the complex and the energies of its constituents ($E_{i}$):

\begin{equation}
\Delta E=E-\sum_{i}E_{i}.
\end{equation}

The basis set superposition error (BSSE) is corrected using the counterpoise procedure~\cite{boys1970calculation}. In the case of the optimized complexes within PCM, the {values} of counterpoise correction from the same complexes without water solvent {are} used. Deformation energy
defines the change in geometry between the isolated molecule ($E_{i}^{isolated}$) and molecule within the complex ($E_{i}^{complex}$):

\begin{equation}
E_{def}=E_{i}^{complex}-E_{i}^{isolated}.
\end{equation}

Consequently, the total interaction energy in the molecular complex can be calculated as

\begin{equation}
E_{tot}=\Delta E+E_{def}.
\end{equation}

\section{Calculation Results}
\label{results}
{Structural parameters of the individual molecules considered in the present work} are taken from~\cite{Saenger} for {the MM} method and are obtained by geometry optimization in the framework of the quantum-chemistry approach. {The} H$_2$O$_2$ molecule is symmetric and can be characterized by the distances between two oxygen atoms, between oxygen and hydrogen atoms, by the angle O-O-H and the dihedral angle H-O-O-H. Tabl. \ref{tab:geometries} shows that the geometrical parameters which were chosen for the calculations in the {MM} method, as well {as} the values obtained by different methods of quantum-chemical approach, are very similar. Also{,} they are comparable with the values obtained in the work~\cite{gonzalez1997}. Note that the dihedral angle H-O-O-H can be very sensitive to the environment. Thus, this value can be considerably different for the isolated molecule and the molecule within a complex. The geometry of {the} H$_2$O$_2$ molecule in the water solution (PCM model) is very similar to its geometry in the gas phase, but with significant differences in their dihedral angles (Tabl. \ref{tab:geometries}). Moreover{,} as {a} water molecule has no {conformational degree of freedom}, its structure is not {as} sensitive to the environment as the structure of {an} H$_2$O$_2$ molecule.

\begin{table*}

\caption{The geometries of H$_{2}$O$_{2}$ and H$_{2}$O molecules that are used for the calculations in the framework of {the MM} method and that are optimized using the corresponding methods of {the} quantum-chemical approach. Distances are given in {\AA}, angles in degrees.}
\label{tab:geometries}
\begin{center}
\begin{tabular}{llllllllll}
\hline\noalign{\smallskip}
\multirow{3}{*}{Solvent} & \multirow{3}{*}{Method} & \ \ & \multicolumn{4}{c}{\underline{H$_{2}$O$_{2}$}} & \ \ \ & \multicolumn{2}{c}{\underline{H$_{2}$O}} \\
 &  & \ \ & O-O & H-O & $\angle$ O-O-H & dihedral & \ \ \ & H-O & $\angle$ H-O-H \\
 \noalign{\smallskip}\hline\noalign{\smallskip}

\multirow{3}{*}{Gas phase} & HF & \ \ & 1.39 & 0.94 & 102.9 & 117.1 & \ \ \ & 0.94 & 106.2 \\
 & B3LYP & \ \ & 1.45 & 0.97 & 100.5 & 121.1 & \ \ \ & 0.96 & 105.0 \\
 & MP2 & \ \ & 1.45 & 0.96 & 99.6 & 121.0 & \ \ \ & 0.96 & 103.5 \\
\noalign{\smallskip}\hline\noalign{\smallskip}

\multirow{3}{*}{PCM} & HF & \ \ & 1.38 & 0.95 & 103.9 & 101.6 & \ \ \ & 0.94 & 105.6 \\
 & B3LYP & \ \ & 1.45 & 0.97 & 101.6 & 105.7 & \ \ \ & 0.96 & 104.5 \\
 & MP2 & \ \ & 1.44 & 0.97 & 100.8 & 107.1 & \ \ \ & 0.96 & 103.1 \\
\noalign{\smallskip}\hline\noalign{\smallskip}
\multicolumn{2}{c}{{MM}} & \ \ & 1.47 & 0.96 & 94.78 & 111.6 & \ \ \ & 0.96 & 106.0 \\
\noalign{\smallskip}\hline

\end{tabular}

\end{center}
\end{table*}

\begin{figure}
\begin{center}

\resizebox{0.3\textwidth}{!}
{%
  \includegraphics{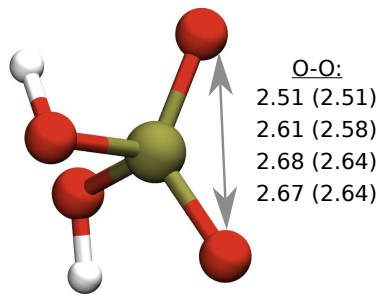}
}
\caption{Spatial structure of {the} PO$_4^-$ group. {The} distances are {listed for} the different calculation methods in the following order (from top to bottom): {MMv}, HF, B3LYP, MP2. {The values that take} into account {the} implicit solvent ({MMh} and PCM model for the corresponding methods) are shown in parenthesis. {The} distance values are given in {\AA}.}
\label{fig:PO4}       
\end{center}
\end{figure}

The optimized structure of the phosphate group obtained in the quantum-chemical approach is more complicated. In this regard, we present only the distance between the two oxygen atoms, which contain the negative charge (Fig. \ref{fig:PO4}). The total charge on {the} phosphate group is $-e$ ($e$ {is the elementary charge}). In the framework of the quantum-chemical approach, the phosphate group PO$_4^-$ is considered as a part of {a} DNA backbone with two hydrogen atoms placed instead of the backbone atoms.

{First of all}, the geometry optimization of the PO$_4^-$ group was made. {Next,} this geometry was {kept} fixed. In the calculations of molecular complexes{, the} PO$_4^-$ group is considered as {a} rigid structure. In the framework of {the MM method,} the interaction of H$_2$O$_2$ and H$_2$O molecules {with only} two oxygen atoms that are open to the solvent is considered, {while the} other atoms of the PO$_4^-$ group {were not taken into account}. The charge on each of these oxygen atoms is considered to be equal to $-0,5|e|$ {(Tabl. S1, Supp. Mat.)}.

\subsection{Complexes consisting of H$_2$O$_2$ and H$_2$O molecules with the phosphate group}
\label{sec:2}

\begin{figure*}[h]
\begin{center}

\resizebox{1\textwidth}{!}{%
  \includegraphics{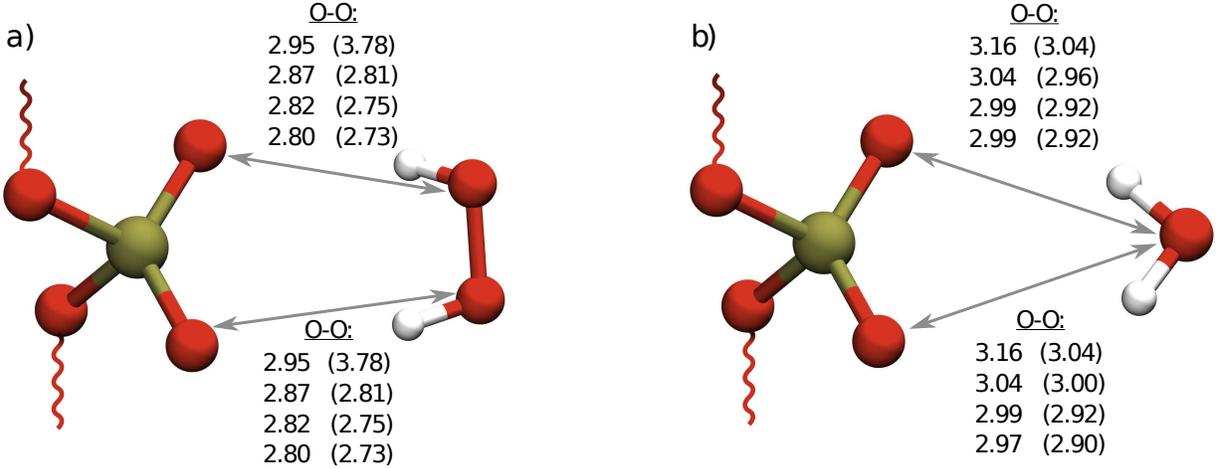}
}
\caption{{Spatial structures of the} complexes consisting of H$_2$O$_2$ (a) and H$_2$O (b) molecules with {the} PO$_4^-$ group. To simplify the visualization, imaginary connections to the DNA backbone are indicated by wavy lines. Distances (in {\AA}) are obtained from the different calculation methods in the following order (from top to bottom): {MMv}, HF, B3LYP, MP2.  {The} values {that take} into account {the} implicit solvent ({MMh} and PCM model for the corresponding methods) are shown in parenthesis. Arrows indicate the distances between the heavy atoms.}
\label{fig:H2O2H2OPO4}       
\end{center}
\end{figure*}

Let us {consider} the complexes of hydrogen peroxide and water molecules with the phosphate group. {The optimized geometries of the complexes} are shown in Fig. \ref{fig:H2O2H2OPO4}. As PO$_4^-$ group is considered as a part of DNA backbone, we take into account only those complexes, where H$_2$O$_2$ and H$_2$O molecu\-les are situated near {the} two oxygen atoms which are open into solution (on the right side of PO$_4^-$ on Fig. \ref{fig:H2O2H2OPO4}).

Fig. \ref{fig:H2O2H2OPO4} shows that hydrogen peroxide and water molecu\-les are situated almost symmetrically {with respect to} the phosphate group and form hydrogen bonds with the charged oxygen atoms of the PO$_4^-$. Due to the {geometric} inequalities between the H$_2$O$_2$ and H$_2$O molecules, the O...O distances in these complexes are different. In the complexes with the hydrogen peroxide molecule (Fig. \ref{fig:H2O2H2OPO4} a) the H-bonds are more straight, thus{,} the O...O distances are smaller. In the case of hydrogen peroxide{,} there are two hydrogen bonds. In the framework of quantum-chemistry approach, hydrogen peroxide molecule is {deformed} {within all the methods used in the present work}. Its dihedral angle is close to 55$^\circ$ which is lower than the corresponding value in the isolated molecule (Tabl. \ref{tab:geometries}). Such a change in the dihedral angle makes a contribution into the deformation energy (about 3-4 kcal/mol for different methods). At the same time, {the corresponding hydrogen bonds are bent significantly in the complex with the water molecule} (angles O-H-O are near 145$^\circ$). {This} results in the weakening of the interaction energy (Fig. \ref{fig:H2O2H2OPO4}). Moreover, the energy difference between the complexes with H$_2$O$_2$ and H$_2$O molecules is large enough ($\approx$6 kcal/mol in {the} gas phase and $\approx$4-5 kcal/mol within PCM). In {the MM method,} the energy difference between the corresponding complexes is much lower because the dihedral angle of {the} H$_2$O$_2$ molecule in the framework of this method is rigid. Note, that such {a} significant interaction energy difference was not obtained {in our previous work~\cite{pyatEPJ2015}} due to the rigidity of the dihedral angle of {the} H$_2$O$_2$ molecule.

\begin{table}
\begin{center}
\caption{Interaction energies of the complexes of  H$_{2}$O$_{2}$ and H$_2$O molecules with {the} PO$_{4}^{-}$ group ({the} spatial {structures are} shown in Fig. \ref{fig:H2O2H2OPO4}). Energy values are given in kcal/mol.}
\begin{threeparttable}[t]
\noindent{\footnotesize
\label{tab:H2O2H2OPO4}
\begin{tabular}{llllll}
\hline\noalign{\smallskip}
Solvent & Method & H$_{2}$O$_{2}\bullet$PO$_{4}^{-}$ & H$_{2}$O$\bullet$PO$_{4}^{-}$ \\
\noalign{\smallskip}\hline\noalign{\smallskip}
 & HF &  -20.1 & -14.2 \\
Gas phase & B3LYP & -21.4 & -14.8 \\
 & MP2 & -19.9 & -14.3 \\
\noalign{\smallskip}

 & HF & -7.0  & -3.3 \\
PCM & B3LYP & -9.8  & -4.8 \\
 & MP2 &  -8.1  & -4.0 \\
\noalign{\smallskip}\hline\noalign{\smallskip}
\multicolumn{2}{c}{{MMv}\tnote{1}} & -11.0 & -12.0 \\
\multicolumn{2}{c}{{MMh}} & -5.6 & -4.9\\
\noalign{\smallskip}\hline
\end{tabular}
\begin{tablenotes}\footnotesize

\item [1] calculated in gas phase~\cite{pyatEPJ2015}

\end{tablenotes}
}
\end{threeparttable}
\end{center}
\end{table}

\subsection{Complexes consisting of H$_2$O$_2$ and H$_2$O molecules with {the} phosphate group in the presence of sodium counterion}
\label{sec:3}

The DNA macromolecule in {a} cell nucleus is situated in {a} water-ionic solution~\cite{Saenger}. This means that the DNA phosphate groups are neutralized by alkali metal ions (Na$^+$, K$^+$, Li$^+$). Consequently, the interaction of the solvent molecules with DNA atomic groups can take place in the presence of counterions. Since {the} sodium ion is one of the most {widespread} ones in a living cell, in the present work the interaction with the sodium (Na$^+$) counterion  is taken into account. {In this way}, we will consider the complex of Na$^+$ with PO$_4^-$ group and then determine how the presence of the counterion can influence the interaction of water and hydrogen peroxide molecules with the phosphate group.

\begin{figure}
\begin{center}

\resizebox{0.9\textwidth}{!}{%
  \includegraphics{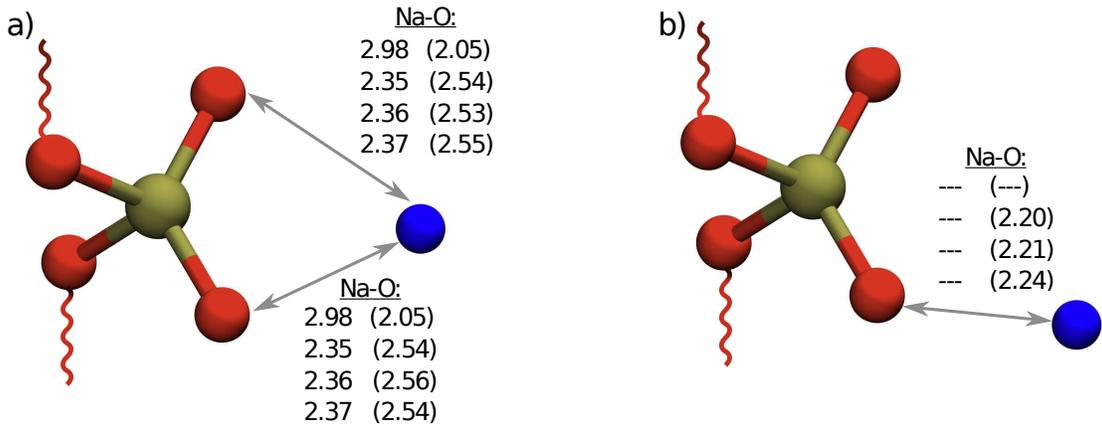}
}
\caption{ Spatial structures of the complexes consisting of {the} PO$_4^-$ group with {the} sodium (Na$^+$) counterion. To simplify the visualization, imaginary connections to the DNA backbone are indicated by wavy lines. Distances (in {\AA}) are obtained from the different calculation methods in the following order (from top to bottom): {MMv}, HF, B3LYP, MP2.  {The values that take} into account {the} implicit solvent ({MMh} and PCM model for the corresponding methods) are shown in parenthesis. Arrows indicate the distances between the heavy atoms.}
\label{fig:NaPO4}       
\end{center}
\end{figure}

The optimized geometries of {the} complexes Na$^+\bullet$PO$_4^-$ are shown in Fig. \ref{fig:NaPO4}. {The} geometry (a) can be obtained by both methods ({MM} and quantum chemistry) as in gas phase as well as in water solution, and geometry (b) is only given by the quantum-chemistry approach within the PCM mo\-del. Fig. \ref{fig:NaPO4} shows the distances between Na$^+$ counterion and oxygen atoms of the phosphate group. Due to the certain values of the parameters for the Born-Mayer potential (\ref{for:bornMayer}), these distances are slightly different for {MM} and {the} quantum chemistry approach. {The} interaction energies of {the} Na$^+\bullet$PO$_4^-$ complex presented in Tabl. \ref{tab:NaPO4} show that taking into account {the} implicit solvent reduces the interaction energies by $\approx$5 times within {MMh} and by $\approx$20 times within PCM.

\begin{table}
\begin{center}
\caption{Interaction energies of the complexes of {a} sodium counterion (Na$^+$) with {the} PO$_{4}^{-}$ group {(Fig. \ref{fig:NaPO4})}. Energy values are given in kcal/mol.}
\begin{threeparttable}[h!]
\noindent{\footnotesize
\label{tab:NaPO4}
\begin{tabular}{lll}
\hline\noalign{\smallskip}
Solvent \ \ \ \ \ & Method \ \ \ \ \ & Na$^+\bullet$PO$_{4}^{-}$ \ \ \ \ \ \\
\noalign{\smallskip}\hline\noalign{\smallskip}

 & HF & -127.5   \\
Gas phase (Fig. \ref{fig:NaPO4}a) & B3LYP & -127.4 \\
 & MP2 & -124.2\\
 &  &    \\

 & HF & -5.5  \\
PCM (Fig. \ref{fig:NaPO4}a) & B3LYP & -5.7 \\
 & MP2 &  -4.2 \\
\noalign{\smallskip}

 & HF & -7.8 \\
PCM (Fig. \ref{fig:NaPO4}b) & B3LYP & -7.6 \\
 & MP2 &  -5.9 \\
\noalign{\smallskip}\hline
\noalign{\smallskip}
\multicolumn{2}{c}{{MMv}\tnote{1} \ (Fig. \ref{fig:NaPO4})}  & -122.0  \\
\multicolumn{2}{c}{{MMh} \ (Fig. \ref{fig:NaPO4})} & -21.8 \\
\noalign{\smallskip}\hline
\end{tabular}
\begin{tablenotes}\footnotesize
\item [1] calculated in gas phase~\cite{pyatEPJ2015}

\end{tablenotes}
}
\end{threeparttable}
\end{center}
\end{table}

\begin{figure}[h!]
\begin{center}

\resizebox{1\textwidth}{!}{%
  \includegraphics{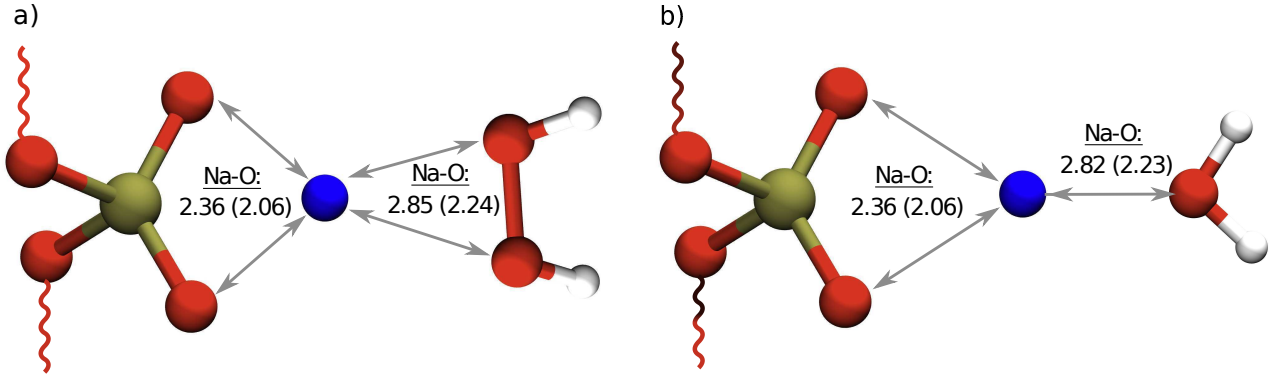}
}
\caption{Spatial structures of {the} complexes consisting of hydrogen peroxide (a) and water {(b)} molecules  with {the} PO$_4^-$ group in the presence of {the} sodium (Na$^+$) counterion obtained by {the MM} method. Distances are given in {\AA}. {The values that take} into account {the} implicit solvent ({MMh}) are shown in parenthesis. Imaginary connections to the DNA backbone are indicated by wavy lines. Arrows indicate the distances between the heavy atoms.}
\label{fig:H2O2H2ONaPO4AAPFh}       
\end{center}
\end{figure}

\begin{figure}
\begin{center}

  \includegraphics{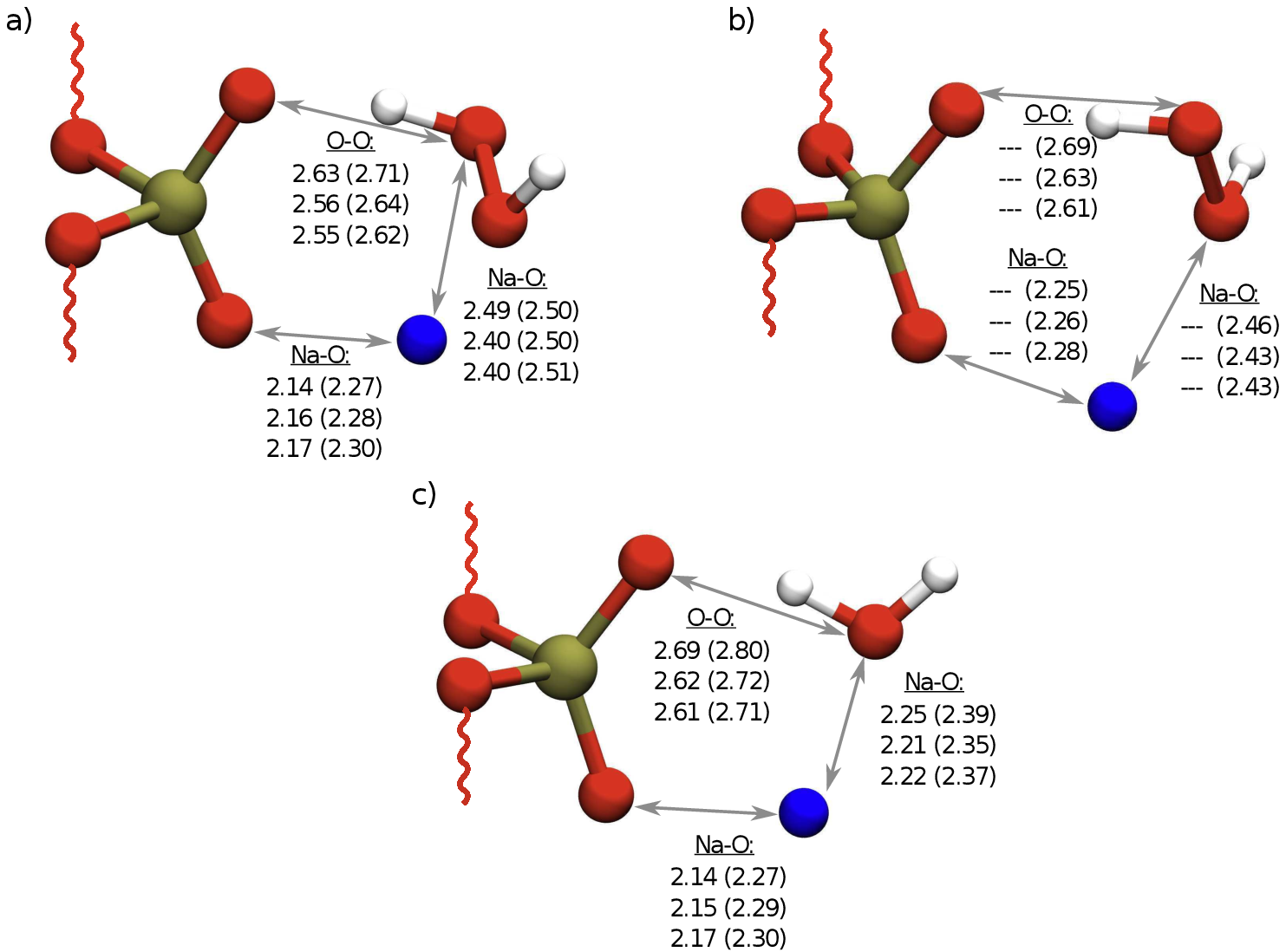}
\caption{Spatial structures of {the} complexes consisting of hydrogen peroxide (a,b) and water (c) molecules with {the} PO$_4^-$ group in the presence of {a} sodium counterion (Na$^+$) obtained by {the} quantum-chemical approach. Distances (in {\AA}) obtained from the different levels of theory are listed in the following order (from top to bottom): HF, B3LYP, MP2. {The values that take} into account {the} implicit solvent (PCM) are shown in parenthesis. Imaginary connections to the DNA backbone are indicated by wavy lines. Arrows indicate the distances between {the} heavy atoms. }
\label{fig:H2O2H2ONaPO4qc}       
\end{center}
\vspace{-0.5cm}
\end{figure}

Let us consider the molecular complexes consisting of three components Na$^+\bullet$H$_2$O$_2\bullet$PO$_4^-$ and Na$^+\bullet$H$_2$O$\bullet$PO$_4^-$. Figs. \ref{fig:H2O2H2ONaPO4AAPFh} and \ref{fig:H2O2H2ONaPO4qc} show that {the MM} method and quantum-che\-mistry approach give different spatial structures. {The} MM method leads to the optimized geometry where {the} H$_2$O$_2$ or H$_2$O molecule is situated symmetrically {with respect} to {the} two oxygen atoms of {the} PO$_4^-$ group with {the} Na$^+$ counterion between them (Fig. \ref{fig:H2O2H2ONaPO4AAPFh} a,b). Tabl. \ref{tab:H2O2H2ONaPO4} shows that for {the} both {MMv} and {MMh} methods the interaction in the complex with hydrogen peroxide is $\approx$ 1 kcal/mol more energetically favourable than in the corresponding complex with {the} water molecule.

{At the same time,} quantum chemistry approach gives the structure where {the} sodium counterion and hydrogen peroxide or water molecule are each situated near the different oxygen atoms of PO$_4^-$ that are open into solution (Fig. \ref{fig:H2O2H2ONaPO4qc} a,c). In other words, the addition of {a} sodium counterion into the two-molecular complexes H$_2$O$_2\bullet$PO$_4^-$ and H$_2$O$\bullet$PO$_4^-$ leads to the displacement of {the} peroxide or water molecule with comparison to the complexes without {the} counterion (Fig. \ref{fig:H2O2H2OPO4} a,b). {The main reason for this displacement is the charge redistribution along the PO$_4^-$. Our results show that the differences between the charges on the oxygen atoms of the phosphate group are $\sim 0.1e$ for various levels of theory. This {charge difference} turns out to be enough to attract {the} Na$^+$ and H$_2$O$_2$ (H$_2$O) towards the different oxygen atoms. The obtained} structures are almost {equal} for the calculations in {the} gas phase and in water solution (PCM model). In both cases{,} H$_2$O$_2$ or H$_2$O molecule forms one hydrogen bond with {the} phosphate group.  Additionally, {the} PCM model gives a structure of the complex with hydrogen peroxide where the O-O distance from H$_2$O$_2$ molecule is situated {almost in the O$^-$-P-O$^-$} plane of the phosphate group (Fig. \ref{fig:H2O2H2ONaPO4qc} b). As can be seen from the Tabl. \ref{tab:H2O2H2ONaPO4}, this structure is $\approx$1 kcal/mol more stable than those shown in Fig. \ref{fig:H2O2H2ONaPO4qc} a.

All the interaction energies for the considered complexes are presented in Tabl. \ref{tab:H2O2H2ONaPO4}. {Taking into account the} PCM model {lowers significantly} the interaction energy but the complex with H$_2$O$_2$ remains to be more energetically {favourable} than the same complex with {the} water molecule. It also should be noted that {the} complexes presented {in} Fig. \ref{fig:H2O2H2ONaPO4qc} c are more probable to be found in crystal structures experimentally~\cite{schneider1996}.

To sum up this part of our study, the addition of {the} sodium counterion to the complexes H$_2$O$_2\bullet$PO$_4^-$ and H$_2$O$\bullet$ PO$_4^-$ significantly influences the interaction of hydrogen peroxide and water molecules with {a} phosphate group. Herewith, the interaction energy in {the} gas phase increases up to $\sim$100 kcal/mol that is very close to the energy of the covalent bond formation. Taking {the} implicit solvent into account makes the obtained values more realistic and comparable with the energy barriers of the intramolecular interactions that take place in {a} living cell~\cite{biochemistry}.

\begin{table}
\begin{center}
\caption{Interaction energies of the complexes of  H$_{2}$O$_{2}$ and H$_2$O molecules with {the} PO$_{4}^{-}$ group in the presence of Na$^+$ ion. Energy values are given in kcal/mol.}
\begin{threeparttable}[h]
\noindent{\footnotesize
\label{tab:H2O2H2ONaPO4}
\begin{tabular}{llll}
\hline\noalign{\smallskip}
Solvent & Method & Na$^+\bullet$H$_{2}$O$_{2}\bullet$PO$_{4}^{-}$ & Na$^+\bullet$H$_{2}$O$\bullet$PO$_{4}^{-}$  \\
\noalign{\smallskip}\hline\noalign{\smallskip}
Gas & HF & -148.8 & -148.2 \\
phase & B3LYP & -150.4 & -149.3 \\
(Fig. \ref{fig:H2O2H2ONaPO4qc} a,c) & MP2 & -145.1 & -144.1 \\
 \noalign{\smallskip}

\multirow{2}{*}{PCM} & HF &    -13.4   & -13.2 \\
\multirow{2}{*}{(Fig. \ref{fig:H2O2H2ONaPO4qc} a,c)} & B3LYP & -15.0  & -14.9 \\
 & MP2 & -11.8 & -11.7 \\
  \noalign{\smallskip}
\multirow{2}{*}{PCM} & HF &    -14.5    & ---\\
\multirow{2}{*}{(Fig. \ref{fig:H2O2H2ONaPO4qc} b)} & B3LYP & -16.4  & ---\\
 & MP2 & -13.2  & ---\\
\noalign{\smallskip}\hline\noalign{\smallskip}

\multicolumn{2}{c}{{MMv}\tnote{1} \ (Fig. \ref{fig:H2O2H2ONaPO4AAPFh})} & -130.3 & -129.2 \\
\multicolumn{2}{c}{{MMh} \ (Fig. \ref{fig:H2O2H2ONaPO4AAPFh})} & -27.4 & -26.5\\

\noalign{\smallskip}\hline
\end{tabular}
\begin{tablenotes}\footnotesize

\item [1] calculated in {the} gas phase~\cite{pyatEPJ2015}

\end{tablenotes}

}
\end{threeparttable}
\end{center}
\end{table}

\section{The possibility of blocking the non-specific DNA recognition sites}
\label{blocking}

{The} calculations performed {in the present work} reveal that hydrogen peroxide molecule can form a complex with {the} PO$_4^-$ group that has {larger} interaction energy as the same complex with {the} water molecule. These results are in accordance with our previous calculations performed by {MMv} method~\cite{pyatEPJ2015}. {The energetical advantage of H$_2$O$_2$ compared to H$_2$O} means that hydrogen peroxide can accumulate near the DNA macromolecule in solution and influence its activity. {Consequently, further processes of genetical information transfer in a living cell such as DNA specific recognition~\cite{ZdorevskyiDopovidi2019} and DNA base pair opening~\cite{ZdorevskyiUjp2019} are much more probable to be blocked.}



{As known, cancer cells differ from healthy ones {in such a way} that in cancer cells DNA replication processes take place more frequently and uncontrollably~\cite{la2017physics}. Thus, the blocking of these processes by hydrogen peroxide must be much more perceptible for cancer cells than for healthy ones. These considerations can explain the `selective' damage by the H$_2$O$_2$ molecules to cancer cells~\cite{chen2005}.}

{Note, that compared with the commonly known mechanism of double-strand breaks, the proposed mechanism of DNA deactivation by blocking its recognition sites, does not imply the rupture of chemical bonds. Therefore, this mechanism can be less sensitive to the DNA repair processes in {a} living cell.}

\section{Conclusions}
\label{conclusions}
In the present work{, the} complexes consisting of {a} hydrogen peroxide or {a} water molecule, phosphate group and {a} sodium counterion are analyzed. Stabilization energies of the considered complexes are calculated using {the molecular mechanics} and quantum-chemistry app\-roach. {The} calculations are performed in {the} gas phase as well as taking into account {the} implicit solvent.  Both methods show that {the} hydrogen peroxide molecule can form a stable complex with DNA phosphate group which is more energetically favourable than {the same complex with the} water molecule. This energetical advantage takes place due to the flexibility of the dihedral angle of {the} hydrogen peroxide molecule. Addition of {a} so\-dium counterion to these complexes makes these interactions much more stable. 

As hydrogen peroxide mole\-cules occur in high concentrations under {heavy-ion irradiation~\cite{Durante2018}}, the formation of such complexes can block the genetical activity of DNA macromolecule of cancer cells and can be an important factor in ion beam therapy treatment.
%
%
\section{Authors contributions}
All the authors were involved in the preparation of the manuscript. All the authors have read and approved the final manuscript.
\section{Acknowledgements}
The present work was partially supported by the Program of Fundamental Research of the Department of Physics and Astronomy of the National Academy of Sciences of Ukraine (project number 0120U100858).

%

%

\end{document}